\documentclass[letterpaper, superscriptaddress, twocolumn, showpacs,
nofootinbib]{revtex4}

\usepackage[]{graphicx}
\usepackage{amsmath}
\usepackage{amsfonts}
\usepackage{color}
\usepackage{hyperref}
\usepackage{bm}
\usepackage{graphicx}
\usepackage{amsbsy}
\usepackage{epstopdf}

\def\ket#1{| #1 \rangle}
\def\bra#1{\langle #1 |}

\def\bk#1#2{\langle #1 | #2 \rangle}

\newcommand{\csign}{\textsc{csign} }
\newcommand\etal{{\em et al.}}

\pacs{03.67.Lx, 03.67.Mn, 03.67.Pp}
 
\date{\today}

\begin{document}

\title{Coherent State Topological Cluster State Production}

\author{C.~R.~Myers}\email{myers@physics.uq.edu.au}
\affiliation{Centre for  Quantum Computation and Communication Technology, School of Mathematics and Physics,
The University of Queensland, St. Lucia 4072 QLD Australia}
\author{T.~C.~Ralph}
\affiliation{Centre for  Quantum Computation and Communication Technology, School of Mathematics and Physics,
The University of Queensland, St. Lucia 4072 QLD Australia}

\begin{abstract}
We present results illustrating the construction of 3D topological cluster states with coherent state logic. Such a construction would be ideally suited to wave-guide implementations of quantum optical processing. We investigate the use of a ballistic \csign gate, showing that given large enough initial cat states, it is possible to build large 3D cluster states. We model $X$ and $Z$ basis measurements by displaced photon number detections and $x$-quadrature homodyne detections, respectively. We investigate whether teleportation can aid cluster state construction and whether the introduction of located loss errors fits within the topological cluster state framework. 
\end{abstract}

\maketitle

\section{\label{sec:intro} Introduction}

The recent use  of topological techniques in the cluster state quantum computation scheme~\cite{Raussendorf07a, Raussendorf07b} has lead to some very encouraging fault tolerant threshold predictions. The initial estimate for the computational basis error threshold was $0.75\%$~\cite{Raussendorf07a, Raussendorf07b}, but it is believed this could reach as high as $1\%$~\cite{Fowler09a}, making this architecture a serious contender for scalable quantum computing~\cite{Devitt09a}. This scheme encodes qubits as defects in a three dimensional cluster state, constructing the necessary qubit gates via topological operations on the surface of the code. A remarkable aspect of these 3D topological cluster states is their resilience to loss, with estimates that they could tolerate a located loss error as high as $25\%$~\cite{Barrett10a}, and may even be able to recover from simultaneous computational basis and located loss errors~\cite{Barrett10a}. 

The bare requirements a physical system must posses in order to implement  topological cluster state computation are~\cite{Fowler09a}:  (i) state preparations of $\ket{0}_{L}$ and $\ket{0}_{L}+e^{i\theta}\ket{1}_{L}$; (ii) $X$ and $Z$ basis measurements; (iii) the \csign entangling gate. 
An implementation with optical systems has been proposed~\cite{Devitt09a} based on integrated optics and in-line non-linear elements in the form of single atom cavity electro-dynamics, showing that with so called photonic modules, large scale topological cluster state computation could in principle be implemented. 
An alternate approach is to move all non-linearities off-line, using techniques of linear optical quantum computing (LOQC)~\cite{KLM}. Unfortunately LOQC with photonic qubits involves massive amounts of resource recycling, which would hugely complicate the optical circuit. Surprisingly, an alternate LOQC scheme based on coherent-state logic exists~\cite{Ralph02a, Ralph03a} which requires no recycling of resource states. 
In this paper, we consider the construction of 3D topological cluster states using such a coherent state logic, building on a ballistic type of linear optical \csign gate~\cite{Ralph02a}. 
The ballistic nature of this scheme makes it an ideal candidate for implementation with integrated linear optical quantum circuits~\cite{Politi08a, Matthews09a, Marshall09a, Smith09a}, since, as we will show, given the appropriate integrated optical wave-guide, this scheme is only in principle limited by the construction of the initial qubits.

The use of coherent states for universal quantum computation was first proposed by Ralph~\etal\cite{Ralph02a, Ralph03a}, where qubits were encoded as $\ket{0}_{L}=\ket{-\alpha}$ and $\ket{1}_{L}=\ket{\alpha}$, requiring Bell state measurements for teleportation and a resource of cat states of the form $(\ket{-\alpha}+\ket{\alpha})/\sqrt{2}$. Since the qubits were only approximately orthogonal, with $|\bk{\alpha}{-\alpha}|^{2}=e^{-4\alpha^{2}}$, sufficiently large cat states were required for this scheme to be viable. 
However, the construction of large optical cat states in the laboratory is difficult, the largest cat states created to date via ancilla assisted two photon subtracted squeezed vacuum~\cite{Takahashi08a} have an average number of photons of $1.4^{2}\approx 1.96$ with a fidelity of $0.60$, while initial experiments~\cite{Gerrits10a} with three photon subtracted squeezed vacuum suggest that cat states with an average photon number as high as $1.76^{2}\approx 3.1$ could be produced. Though, even with the hurdle of constructing large amplitude cat states, quantum computation with coherent state logic may still be competitive with other optical quantum computation schemes, since the success probability for the basic gates is quite high and coherent state qubit teleportation is deterministic. A recent extension to this coherent state computation scheme by Lund~\etal~\cite{Lund08a} showed that universal quantum computation was still possible with small amplitude coherent states. 

In this paper, we address the second two physical requirements for three dimensional cluster state production with coherent states. In contrast to~\cite{Ralph03a}, we define logical qubits as $\ket{0}_{L}=\ket{\text{vac}}\equiv \ket{0}$ and $\ket{1}_{L}=\ket{\alpha}$~\cite{Ralph02a}, both definitions being equivalent up to a displacement in phase space. As in~\cite{Ralph02a, Ralph03a}, we are confined by the inherent error associated with our non-orthogonal qubit definition, with $\bk{0}{\alpha}=e^{-|\alpha|^{2}/2}$, and the feasibility of constructing large amplitude cat states of the form $\ket{0}+\ket{\alpha}$, the basic states required for cluster state production in this scheme. 
In Section~\ref{sec:basicscheme}, we use a beam splitter as our basic \csign gate~\cite{Ralph02a}, showing that it is in principle possible to ballistically construct large 3D cluster states with coherent state logic, finding that coherent cat states with an average number of photons $>85$, meaning cat state with amplitudes larger than $9.25$ ($\ket{\alpha>9.25}+\ket{\alpha<-9.25}$), will result in a computational error rate per qubit $<1\%$. 
Next, in Section~\ref{sec:measurements}, we model $X$ and $Z$ basis measurements by displaced photon number detections and $x$-quadrature homodyne detections, respectively, showing that measurement errors per qubit can be made below $1\%$ for amplitudes above $> 10.42 $ ($\ket{\alpha> 10.42}+\ket{\alpha<-10.42}$), that is, coherent cat states with an average number of photons $>108$. 
In Section~\ref{sec:teleportation}, we make use of teleporation to \emph{clean up} the cluster states built from the basic scheme, showing that we can produce 3D cluster states from low amplitude cat states with arbitrarily high fidelity, albeit at the expense of moving away from the completely ballistic nature of the basic \csign gate and introducing a success probability. 
In Section~\ref{sec:comparison}, we attempt to capitalise on the topological 3D cluster state code's ability to deal with simultaneous computational basis and located loss errors, investigating whether {\em cleaning up} our 3D cluster states with teleportation can reduce the amplitude size of the initial cat state required by trading off computation error for located loss error. We present preliminary results that suggest in the most ambitious scenario for topological cluster states, when $5.35<\alpha<5.54$, it is advantageous to use teleportation. 
Finally, in Section~\ref{sec:conclusions}, we conclude.

\section{\label{sec:basicscheme}Ballisitc Scheme}

\begin{figure}[ht]
\includegraphics[width=6cm]{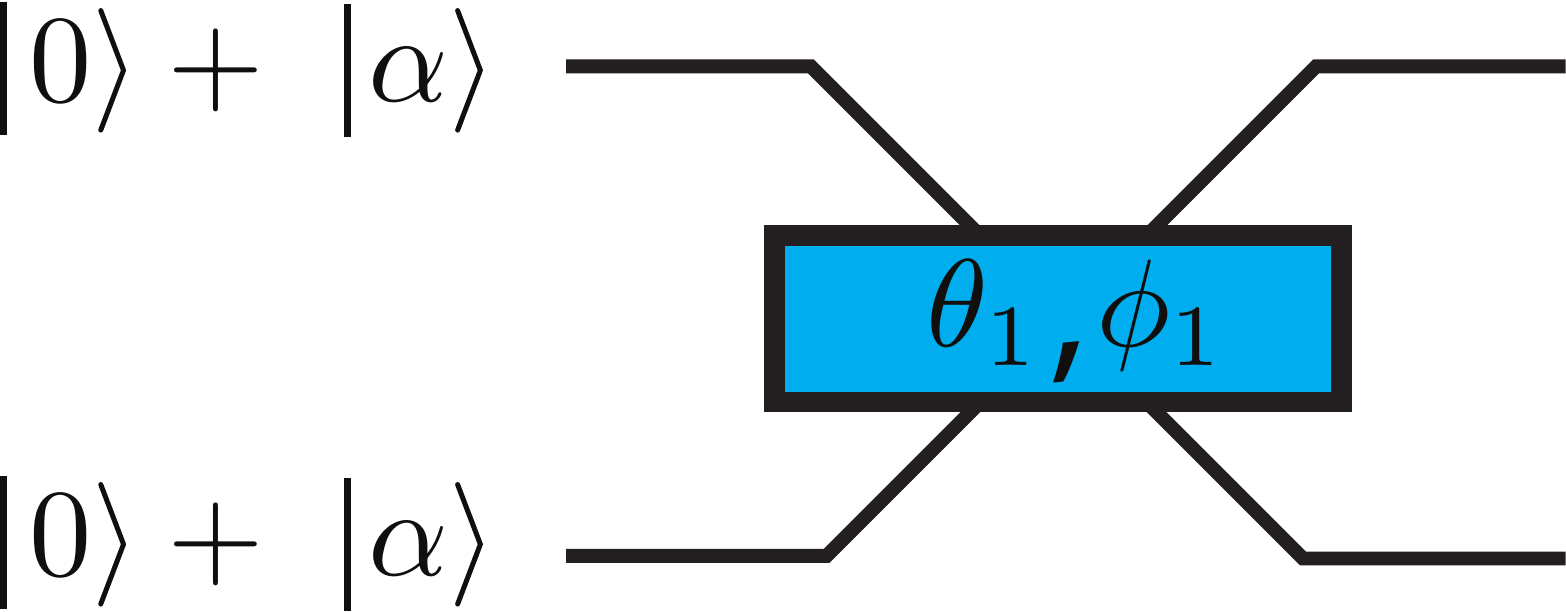}
\caption{\footnotesize Ballisitc \csign gate modelled as a beam splitter of reflectivity $\theta_{1}=\frac{\pi}{2\alpha^{2}}$, $\phi_{1}=-\frac{\pi}{2}$~\cite{Ralph02a}. Note, for clarity, the input cat states are un-normalised.  } 
\label{Fig:CSignBS}
\end{figure}

One of the physical requirements a physical system must posses to be suitable for 3D topological state quantum computing is the ability to perform a \csign gate between neighbouring qubits. We model our basic \csign gate as a symmetric beam splitter of reflectivity of $\theta=\pi/2\alpha^{2}$, $\phi=-\pi/2$~\cite{Ralph02a}, shown in Fig.~(\ref{Fig:CSignBS}). If we consider two cat states of the form 
\begin{align}
\frac{1}{\mathcal{N}}\left(\ket{0}+\ket{\alpha}\right)\otimes \left(\ket{0}+\ket{\alpha}\right),\label{eqn:actualCSstateInitial}
\end{align}
incident on such a beam splitter, the output is given by
\begin{align}
 \frac{1}{\mathcal{N}}\left(\ket{00}\right.&+\ket{i\alpha \sin\theta}\ket{\alpha\cos\theta}\nonumber\\
 &\left.+\ket{\alpha\cos\theta}\ket{i\alpha\sin\theta}+\ket{e^{i\theta}\alpha}\ket{e^{i\theta}\alpha}\right)\label{eqn:actualCSstate},
\end{align}
where $\mathcal{N}=2(1+e^{-\alpha^{2}/2})$ is the normalisation. When $\alpha\gg 1$, Eq.(\ref{eqn:actualCSstate}) is approximately 
\begin{align}
\frac{1}{2}\left(\ket{00}+\ket{0\alpha}+\ket{\alpha 0}-\ket{\alpha\alpha}\right)\label{eqn:idealCSstate}.
\end{align}
This can be seen from the inner product of two coherent states of different amplitude~\cite{WallsBook}:
\begin{align}
\bk{\beta}{\alpha}=\exp\left\{-\frac{1}{2}\left(|\alpha|^{2}+|\beta|^{2}\right)+\alpha\beta^{*}\right\}.
\end{align}
The approximate $\ket{11}_{LL}$ state picks up a $\pi$ phase:
\begin{align}
\bk{\alpha,\alpha}{e^{i\theta}\alpha, e^{i\theta}\alpha}&=\exp\left\{-2\alpha^{2}\left(1-e^{i\theta}\right)\right\}\\
&\approx \exp\left\{2i\theta \alpha^{2}\right\}=-1\nonumber,
\end{align}
while the approximate $\ket{01}_{LL}$ and $\ket{10}_{LL}$ states do not pick up any phase factors: 
\begin{align}
\bk{0,\alpha}{i \alpha\sin\theta, \alpha\cos\theta}&=\exp\left\{\alpha^{2}\left(1-\cos\theta\right)\right\}1,\\
&\approx 1\nonumber.
\end{align}
where, since $\theta=\pi/2\alpha^{2}$ and $\alpha\gg 1$, we can assume $\theta\ll 1$. 

\begin{figure}[ht]
\includegraphics[width=6cm]{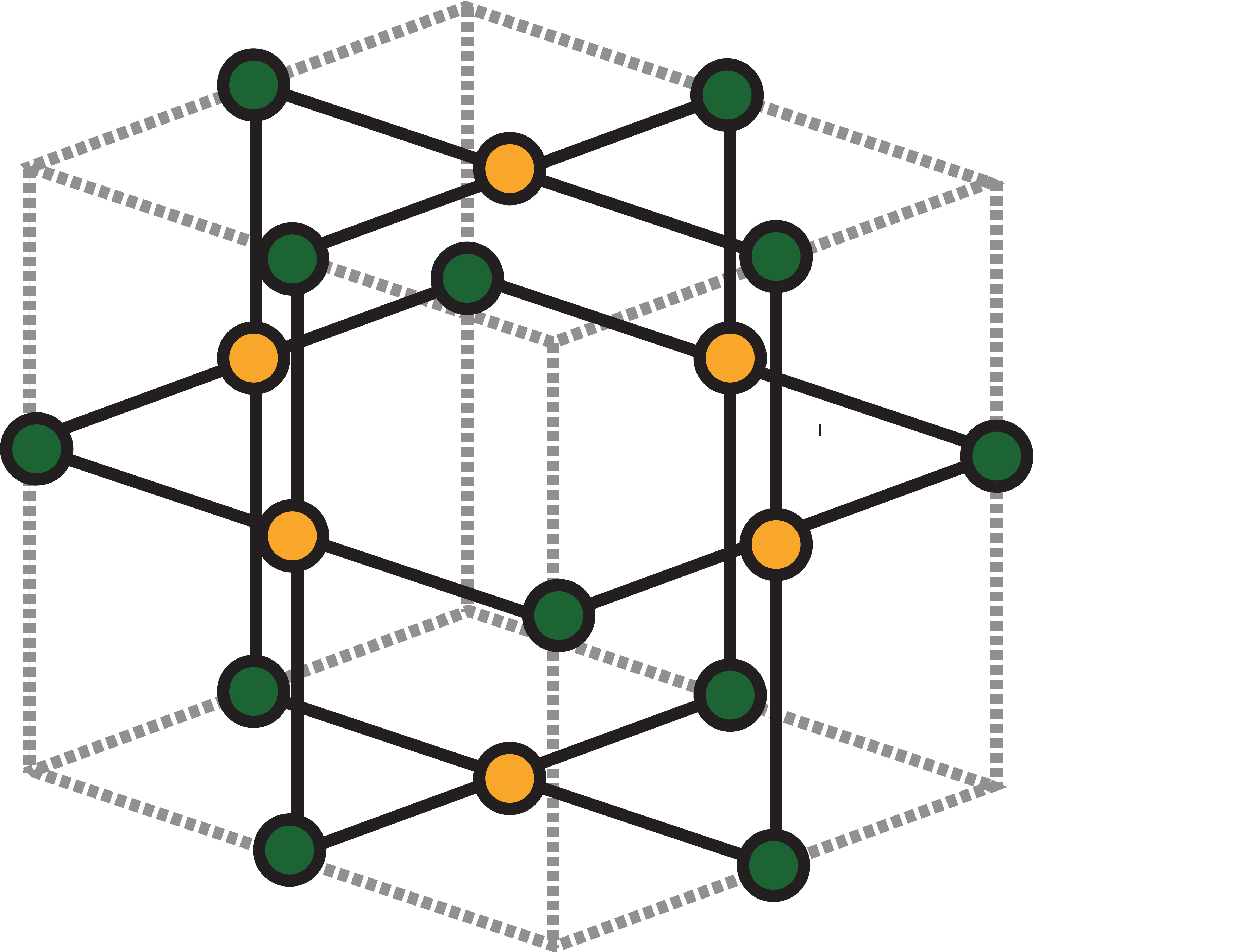}
\caption{\footnotesize Unit cell for the 3D topological cluster state~\cite{Raussendorf07a, Raussendorf07b}. The solid black lines between coloured qubit circles represent \csign gates. } 
\label{Fig:ClusterUnit}
\end{figure}

\begin{figure}[ht]
\includegraphics[width=5cm]{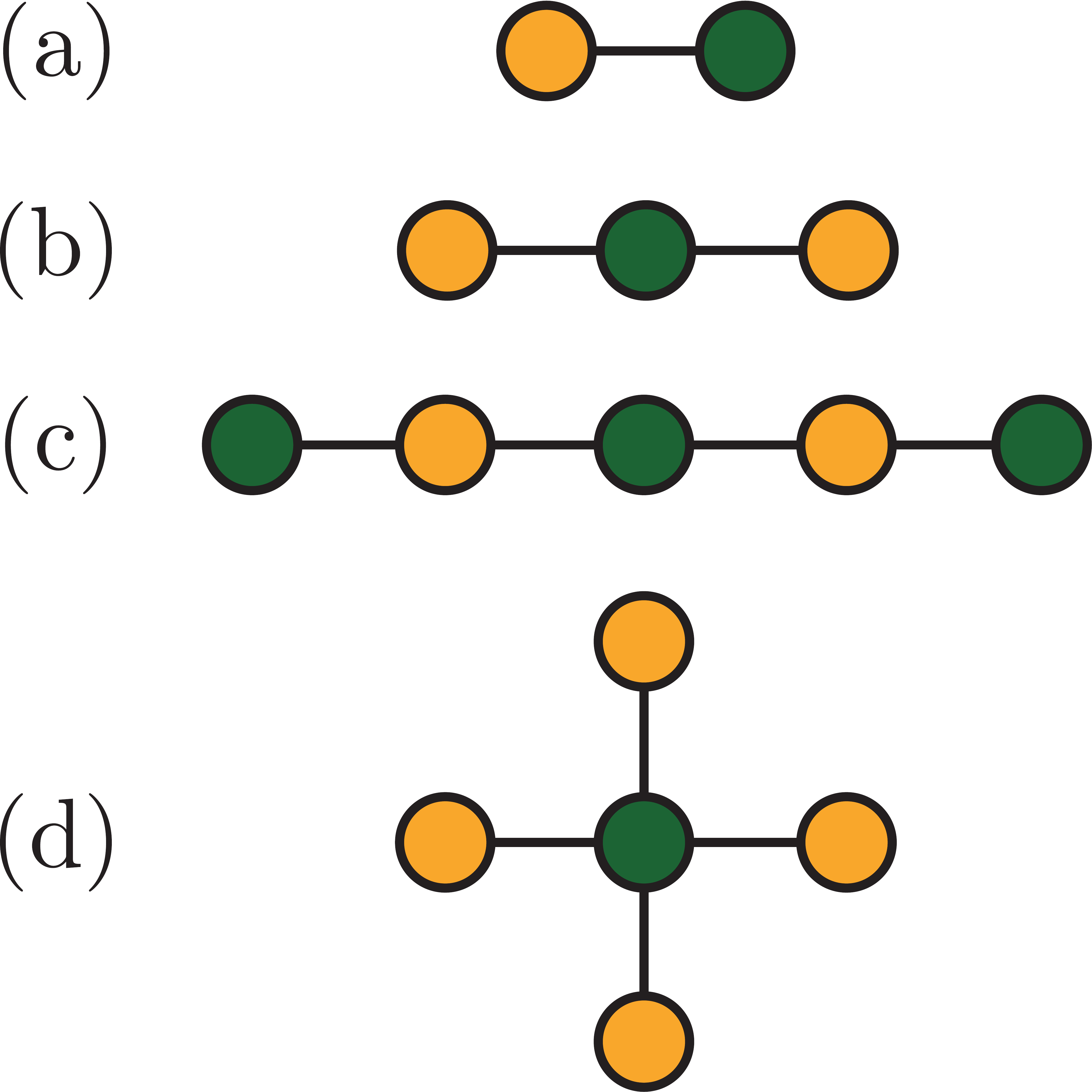}
\caption{\footnotesize The progressive basic building blocks for the 3D topological cluster state unit cell in Fig.~(\ref{Fig:ClusterUnit}):  (a) The 2 qubit cluster; (b) The 3 qubit cluster; (c) The 5 qubit linear cluster; (d) The 5 qubit star cluster. Each vertical/horizontal line corresponds to a \csign gate constructed from the beam splitter shown in Fig.~(\ref{Fig:CSignBS}). Each coloured qubit circle is initially in the cat state $\left(\ket{0}+\ket{\alpha}\right)/\mathcal{N}$. } 
\label{Fig:PictureStates}
\end{figure}

In order to build the 3D topological cluster state with coherent state qubits, we first consider the construction of the elementary building blocks for the cluster state unit cell~\cite{Raussendorf07a, Raussendorf07b}, shown in Fig.~\ref{Fig:ClusterUnit}. These progressive building blocks are shown in Fig.~(\ref{Fig:PictureStates}). 

\begin{figure}[ht!]
\includegraphics[width=8cm]{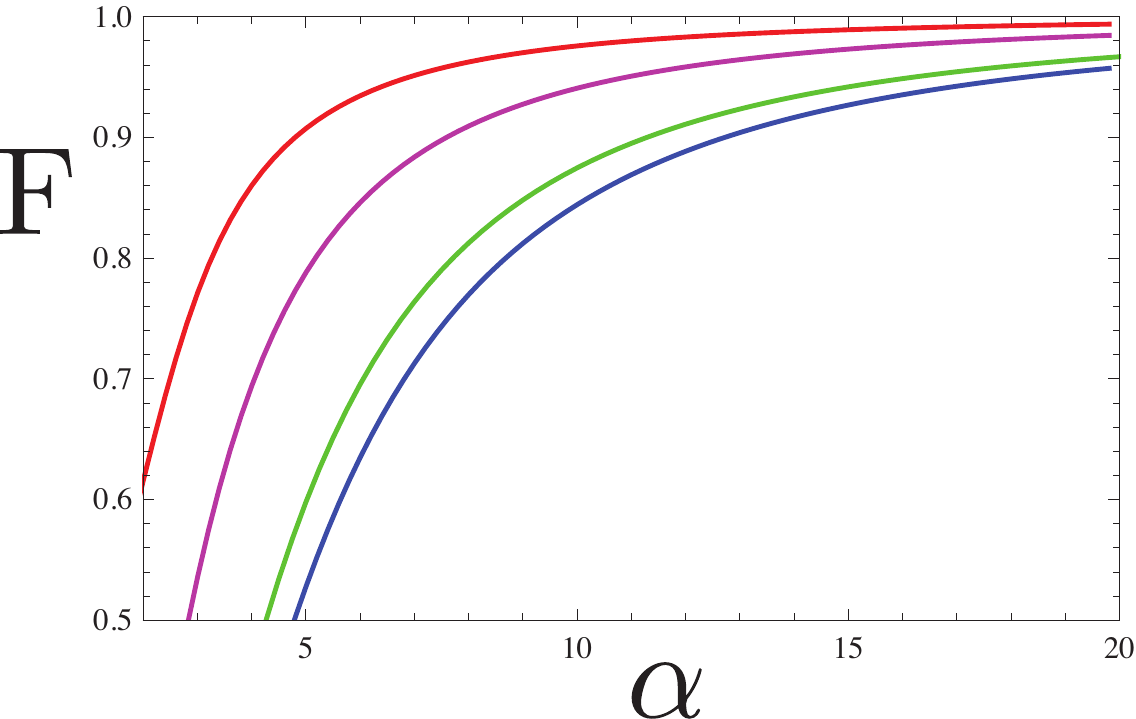}
\caption{\footnotesize The fidelity of constructing the progressive basic building blocks from Fig~(\ref{Fig:PictureStates}) with the beam splitter \csign gate in Fig.~(\ref{Fig:CSignBS}). In \textcolor{red}{red} the two qubit cluster (Fig.~(\ref{Fig:PictureStates})a) fidelity is shown; in \textcolor{magenta}{magenta} the three qubit cluster (Fig.~(\ref{Fig:PictureStates})b) fidelity is shown;  in \textcolor{green}{green} the five qubit linear  cluster (Fig.~(\ref{Fig:PictureStates})c) fidelity is shown; in \textcolor{blue}{blue} the five qubit star cluster (Fig.~(\ref{Fig:PictureStates})d) fidelity is shown. }
\label{Fig:BasicFidelitiesPlot}
\end{figure}

To calculate the fidelity for each of the cluster states in Fig.~(\ref{Fig:PictureStates}), we compare to a hypothetical cluster state made with ideal coherent state \csign gates of the from given in Eq.~(\ref{eqn:idealCSstate}). These fidelities are shown in Fig.~(\ref{Fig:BasicFidelitiesPlot}). Notice that as the amplitude size of the initial cat state increases, the fidelity approaches 1. Also, as expected, the fidelity decreases as the complexity of the cluster state increases. 

\begin{figure}[ht!]
\includegraphics[width=8cm]{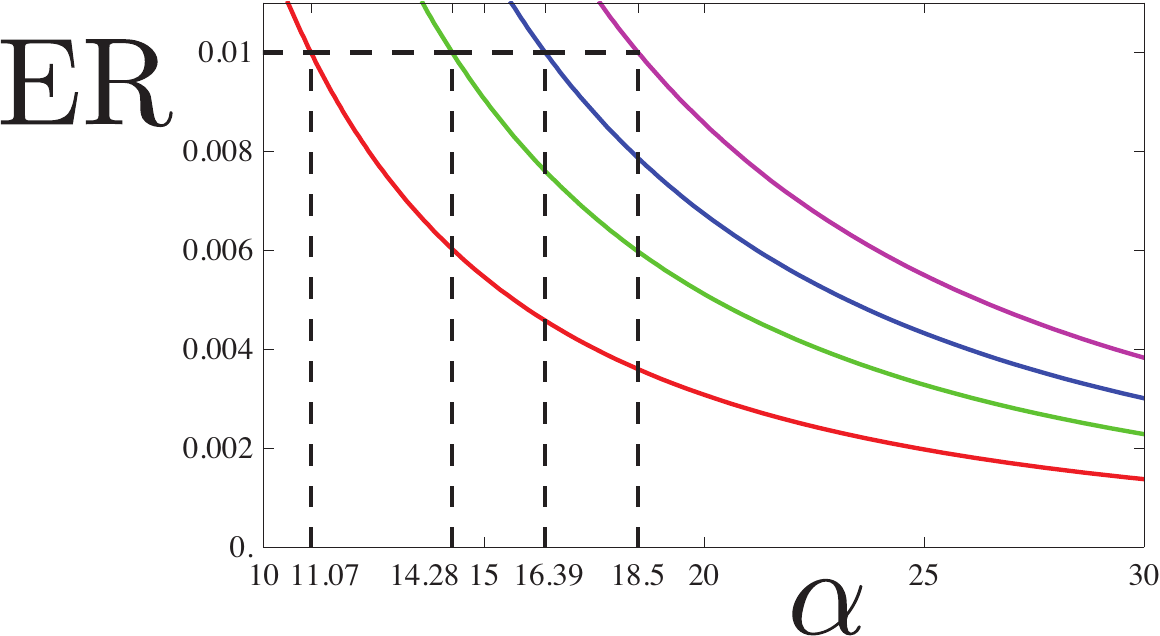}
\caption{\footnotesize The error rate per qubit (ER) for each of the unit cell building blocks in Fig.~(\ref{Fig:PictureStates}). In \textcolor{red}{red} the two qubit cluster (Fig.~(\ref{Fig:PictureStates})a) ER is shown; in \textcolor{green}{green} the three qubit cluster (Fig.~(\ref{Fig:PictureStates})b) ER is shown;  in \textcolor{blue}{blue} the five qubit linear  cluster (Fig.~(\ref{Fig:PictureStates})c) ER is shown; in \textcolor{magenta}{magenta} the five qubit star cluster (Fig.~(\ref{Fig:PictureStates})d) ER is shown. The $\alpha$ values for each cluster state that gives an ER of $1\%$ are also shown.}
\label{Fig:BasicFidelitiesPlotER}
\end{figure}

We gauge the capability of the ballistic \csign gate to construct 3D cluster states by calculating the error rate per qubit (ER) for each of the building block cluster states in Fig.~(\ref{Fig:PictureStates}) by comparing the fidelities in Fig.~(\ref{Fig:BasicFidelitiesPlot}) with the fidelities from an ideally constructed qubit cluster state in which each qubit has undergone a depolarising error~\cite{NielsenBook}, 
as shown in Fig.~(\ref{Fig:BasicFidelitiesPlotER}). 
Note that when $\alpha=11.07$, that is, when the initial cat states are of the form $\ket{\alpha=0}+\ket{\alpha=11.07}$, the error rate per qubit for the two qubit cluster is below $1\%$, the expected maximum computation basis error that the  3D topological cluster state code could correct~\cite{Fowler09a}. When $\alpha=18.5$, the error rate per qubit for the five qubit star cluster in Fig.~(\ref{Fig:PictureStates}d) is below $1\%$.

\section{\label{sec:measurements}Measurements}

Another of the physical requirements a physical system must possess to be suitable for 3D topological cluster state quantum computing is the ability to measure qubits in the $X$ and $Z$ basis. 

We model $X$ basis measurements by first displacing our state by $-\frac{\alpha}{2}$, followed by a photon number measurement, effectively implementing a photon number parity measurement. That is, given the phase resulting from displacing a coherent state and the projection of a coherent state into the number basis~\cite{WallsBook},
\begin{align}
D(\alpha)D(\beta)&=\exp\left\{i\,\text{Im}[\alpha\beta^{*}]\right\}D\left(\alpha+\beta\right)\\
\bk{n}{\beta}&=\exp\left\{-\frac{|\beta^{2}|}{2}\right\}\frac{\beta^{n}}{\sqrt{n!}}\nonumber,
\end{align}
we find $\bra{n}D(-\frac{\alpha}{2})\ket{\beta}$:
\begin{align}
&=\bra{n}\exp\left\{-\frac{i\alpha}{2}\text{Im}[\beta^{*}]\right\}\ket{\beta-\frac{\alpha}{2}}\nonumber\\
&=\exp\left\{ -\frac{i\alpha}{2}\text{Im}[\beta^{*}]-\frac{1}{2}|\beta|^{2}+\frac{\alpha}{2}\text{Re}[\beta]-\frac{\alpha^{2}}{8}\right\}\frac{\left(\beta-\frac{\alpha}{2}\right)^{n}}{\sqrt{n!}}.\label{Eqn:XbasisMeas}
\end{align}
We gain some insight into how Eq.~(\ref{Eqn:XbasisMeas}) models an $X$ basis measurement by considering the $X$ basis measurement of an arbitrary qubit state $c_{0}\ket{0}+c_{1}\ket{1}=\frac{1}{\sqrt{2}}(c_{0}+c_{1})\ket{X^{+}}+\frac{1}{\sqrt{2}}(c_{0}-c_{1})\ket{X^{-}}$. In terms of coherent state logic, measuring the corresponding state gives
\begin{align}
\bra{X\,\text{basis meas.}}\left(c_{0}\ket{0}+c_{1}\ket{\alpha}\right)=\frac{e^{-\frac{\alpha^{2}}{2}}\left(\frac{\alpha}{2}\right)^{n}}{\sqrt{n!}}\left[(-1)^{n}c_{0}+c_{1}\right].
\end{align}
When $n$ is even, we detect $\ket{X^{+}}$, when  $n$ is odd, we detect $\ket{X^{-}}$. 

\begin{figure}[ht]
\includegraphics[width=6cm]{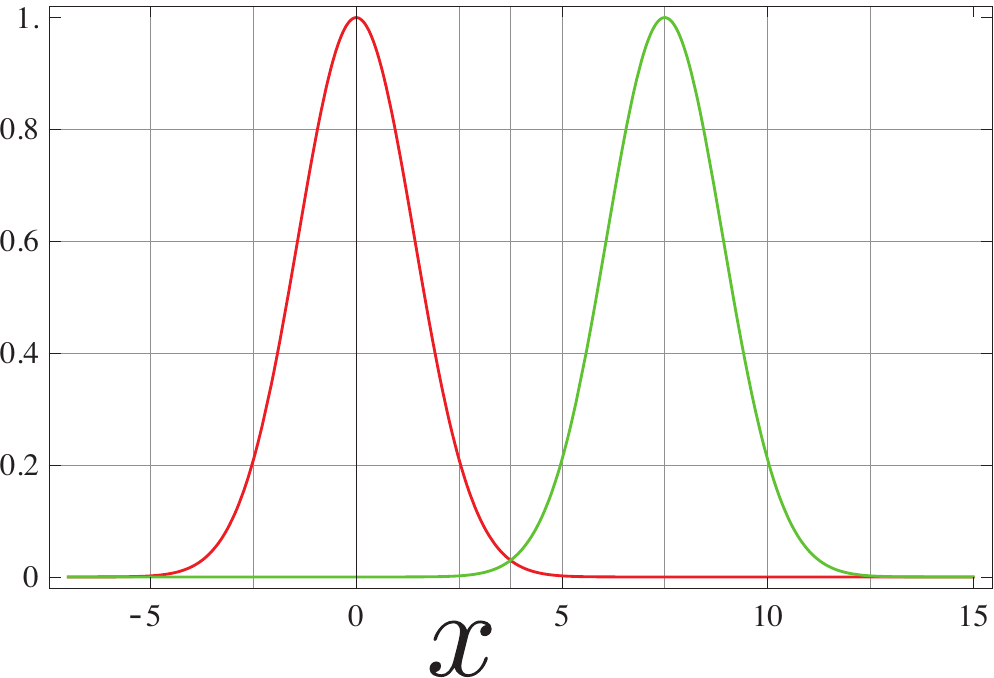}
\caption{\footnotesize The measurement outcome for the detection of $c_{0}\ket{0}+c_{1}\ket{\alpha}$ in the $Z$ basis. In \textcolor{red}{red} we plot $\bk{x}{0}$ and in \textcolor{green}{green} we plot  $\bk{x}{\alpha=3.75}$. } 
\label{Fig:ZHomoDetPlot}
\end{figure}

We model $Z$ basis measurements by homodyne detection in the $x$ quadrature: $\bk{x}{\beta}$. Since the basic \csign gate transforms a small portion of the initially real cat states into the complex direction, as can be seen in Eq.~(\ref{eqn:actualCSstate}), we need to consider the $x-$projection of a general complex coherent state, keeping track of all possible phase terms. We do this by considering both $\bra{x}\hat{a}\ket{\beta}$ and $\bra{x}\hat{a}^{\dagger}\ket{\beta}$. By solving the ODE resulting from $\bra{x}\hat{a}\ket{\beta}=\beta\bk{x}{\beta}$~\cite{GardinerBook}, we find $\bk{x}{\beta}$
\begin{align}
=\left(2\pi\right)^{-\frac{1}{4}}\exp\left\{i\phi+2i\,\text{Re}[\beta]\text{Im}[\beta]-\left(\text{Im}[\beta]\right)^{2}-\frac{1}{4}\left(x-2\beta\right)^{2}\right\},
\end{align}
where we determine the as yet undetermined phase $\phi$ from $\bra{x}\hat{a}^{\dagger}\ket{\beta}$. Given the $x$-projection of a number state~\cite{GardinerBook} and two properties of Hermite polynomials~\cite{AbramowitzBook}: 

\begin{align}
\bk{x}{n}&=\left(2^{n}n!\right)^{-\frac{1}{2}}\left(2\pi\right)^{-\frac{1}{4}}e^{-\frac{x^{2}}{4}}H_{n}\left(\frac{x}{\sqrt{2}}\right), \nonumber\\
\sum_{n=0}^{\infty}\frac{H_{n}(x)t^{n}}{n!}&=e^{2xt-t^{2}},\\
H_{n+1}(x)&=2xH_{n}(x)-2nH_{n-1}(x)\nonumber, 
\end{align}
we find $\bra{x}\hat{a}^{\dagger}\ket{\beta}$
\begin{align}
=\left(2\pi\right)^{-\frac{1}{4}}\left(x-\beta\right)\exp\left\{-\frac{|\beta|^{2}}{2}-\frac{x^{2}}{4}+x\beta-\frac{\beta^{2}}{2}\right\}. \label{Eqn:XbasisMeasStep}
\end{align}
Next, if we consider the ODE associated with $\bra{x}\hat{a}^{\dagger}\ket{\beta}$ given Eq.~(\ref{Eqn:XbasisMeasStep}), we find $\phi=-\text{Re}[\beta]\text{Im}[\beta]$. The $Z$ basis measurement is then found to be 
\begin{align}
\bk{x}{\beta}&=\left(2\pi\right)^{-\frac{1}{4}}\exp\left\{i \,\text{Re}[\beta]\text{Im}[\beta]-\left(\text{Im}[\beta]\right)^{2}-\frac{1}{4}\left(x-2\beta\right)^{2}\right\}.\label{Eqn:ZbasisMeas}
\end{align}
We gain some insight into how Eq.~(\ref{Eqn:ZbasisMeas}) models an $Z$ basis measurement by considering the $Z$ basis measurement of an arbitrary coherent logic state $c_{0}\ket{0}+c_{1}\ket{\alpha}$:
\begin{align}
\bra{Z\,\text{basis meas.}}&\left(c_{0}\ket{0}+c_{1}\ket{\alpha}\right)\nonumber\\
&=\left(2\pi\right)^{-\frac{1}{4}}\left[c_{0}e^{-\frac{x^{2}}{4}}+c_{1}e^{-\frac{1}{4}(x-2\alpha)^{2}}\right].
\end{align}
When we detect $\ket{Z^{+}}$, we expect a homodyne outcome centred around $x=0$ and when we detect $\ket{Z_{-}}$, we expect a homodyne outcome centred around $x=2\alpha$, as shown in Fig.~(\ref{Fig:ZHomoDetPlot}).

We evaluate the effectiveness of modelling $X$ and $Z$ basis measurements by displaced photon number and $x$-homodyne detections by looking at the visibility~\cite{WallsBook}:
\begin{align}
V=\frac{P_{\text{max}}-P_{\text{min}}}{P_{\text{max}}+P_{\text{min}}}
\end{align} 

Since coherent states do not entangle on beam splitters, it is straight forward to calculate the visibility. For example, consider measuring the two qubit cluster state in Fig.~(\ref{Fig:PictureStates}a) in the $XZ$ basis. The initial state for this cluster state is given by 
\begin{align}
\frac{1}{\mathcal{N}}\sum_{m_{1},m_{2}\in\{ 0,\alpha\}}\ket{m_{1},m_{2}}.
\end{align}
After the beam splitter \csign gate, this state becomes
\begin{align}
\ket{\psi_{\text{1BS}}}=\frac{1}{\mathcal{N}}\sum_{m_{1},m_{2}\in\{ 0,\alpha\}}\ket{p_{1},p_{2}},
\end{align}
where 
\begin{align*}
p_{1}&=m_{1}\cos\theta+im_{2}\sin\theta,\\
p_{2}&=im_{1}\sin\theta+m_{2}\cos\theta.
\end{align*}
The probability of a given measurement is then given by $P_{\text{1BS}}=$ $|\bra{X\text{-basis}}\bra{Z\text{-basis}}\cdot\ket{\psi_{\text{1BS}}}|^{2}$ $\equiv$ $|\bra{n}D(\frac{\alpha}{2})\bra{x}\cdot\ket{\psi_{\text{1BS}}}|^{2}$. By writing the ideal two qubit cluster state with the first mode in the $X$ basis and the second mode in the $Z$ basis: $\frac{1}{2}\left(\ket{X^{+}Z^{+}}+\ket{X^{-}Z^{-}}\right)$, we can calculate $P_{\text{max}}$ and $P_{\text{min}}$ for the two qubit cluster: 
$P_{\text{max}}=$ $\left(\int_{-\infty}^{\alpha}dx\sum_{n\,\,\text{even}}+\int_{\alpha}^{\infty}dx\sum_{n\,\,\text{odd}}\right) P_{\text{1BS}}$ 
and 
$P_{\text{min}}=$ $\left(\int_{-\infty}^{\alpha}dx\sum_{n\,\,\text{odd}}+\int_{\alpha}^{\infty}dx\sum_{n\,\,\text{even}}\right) P_{\text{1BS}}$. The reason we integrate from either $\alpha$ to $\infty$ or $-\infty$ to $\alpha$ is to minimise the errors resulting from having non-orthogonal qubits. That is, the midway point between the $Z$ basis outcomes is $x=\alpha$, as can be seen in Fig.~(\ref{Fig:ZHomoDetPlot}). This method can be extended to calculate the visibility for larger cluster states. 

\begin{figure}[ht]
\includegraphics[width=6cm]{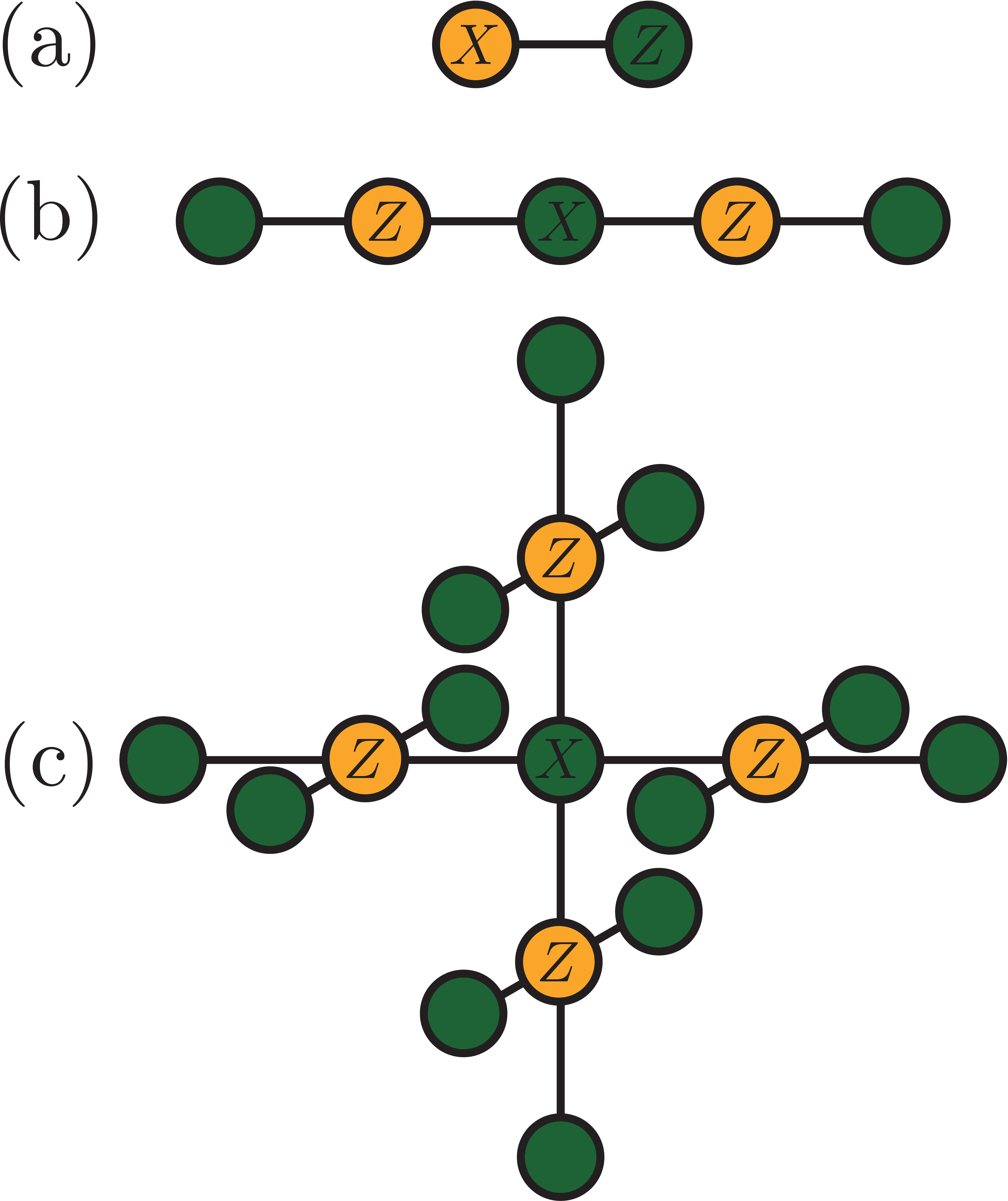}
\caption{\footnotesize Detect $XZ$ for the two qubit cluser. (b) Detect $ZXZ$ for the five qubit linear cluster. (c) Detect the stabiliser $ZZXZZ$ for the seventeen qubit cluster. Each vertical/horizontal line corresponds to a \csign gate.} 
\label{Fig:PictureStatesVis}
\end{figure}

To determine the performance of our measurement model, we measure certain stabilisers of the 3D cluster state. We endeavour to measure the operator $ZZXZZ$, since this is a local stabiliser for the 3D topological cluster state and can also be used to initiate a faulty cluster state into the correct state~\cite{Devitt09a}. We start by measuring smaller operators on the building block cluster states in Fig.~(\ref{Fig:PictureStates}), first measuring the operator $XZ$ on the two qubit cluster, next measuring $ZXZ$ on the five qubit linear cluster and finally measuring $ZZXZZ$ on the maximally connected seventeen qubit star cluster state, as shown in Fig.~(\ref{Fig:PictureStatesVis}). The state shown in Fig.~(\ref{Fig:PictureStatesVis}c) is equivalent to the unit cell in Fig~(\ref{Fig:ClusterUnit}), except in this case we consider the 3D cluster state from the perspective of a single qubit, labelled $X$ in this case, instead of considering the cluster state as a lattice. 

\begin{figure}[ht]
\includegraphics[width=8cm]{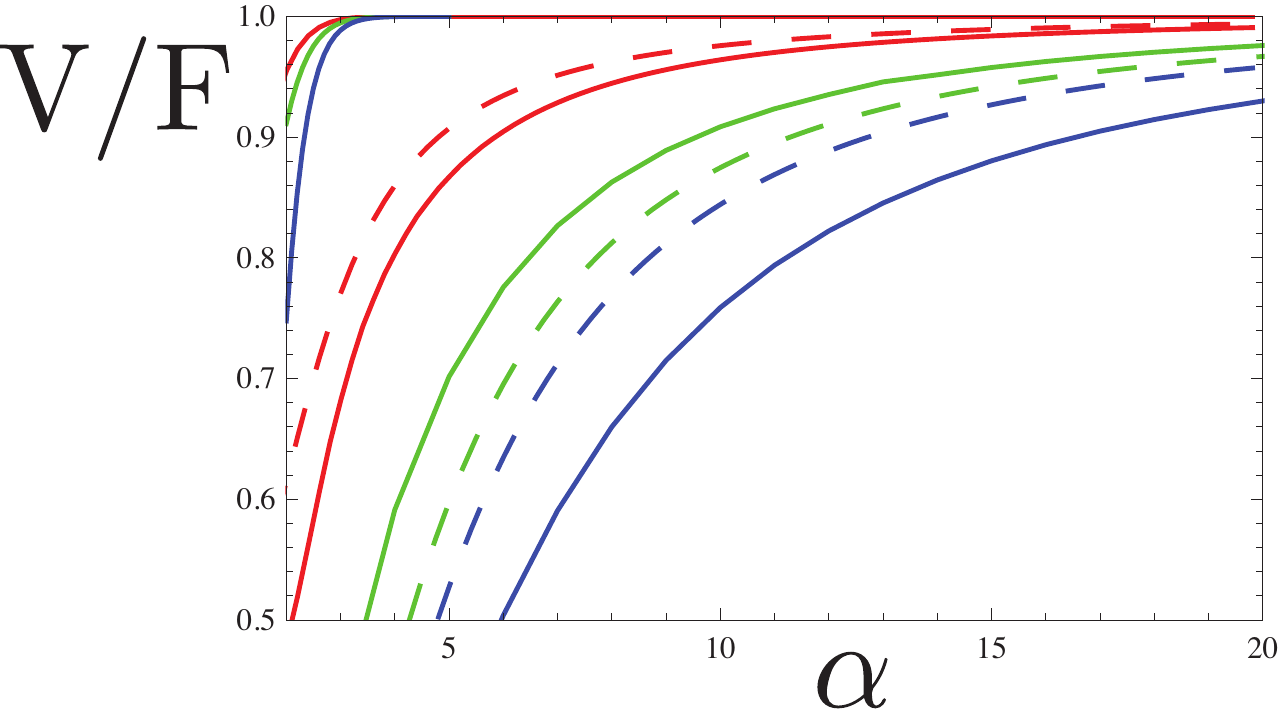}
\caption{\footnotesize The visibility is shown in the solid lines and the fidelity is shown in the dashed lines. The detection of $XZ$ for the two qubit cluster is shown in \textcolor{red}{red}. The detection of $ZXZ$ for the five qubit linear cluster is shown in \textcolor{green}{green}, this line also corresponds to detecting $ZXZ$ for the three qubit cluster. The detection of $ZZXZZ$ for the seventeen qubit cluster is shown in \textcolor{blue}{blue}, this line also corresponds to detecting $ZZXZZ$ for the five qubit star cluster state. In the top left hand corner we show the visibility for the ideal two, five and seventeen qubit cluster states. The fidelity curves correspond to the two qubit cluster, the five qubit linear cluster and the five qubit star cluster.} 
\label{Fig:Vis}
\end{figure}

We plot the visibility as a function of the amplitude size of the initial cat state in Fig.~\ref{Fig:Vis}. It is worth noting that the visibility for measuring $ZXZ$ for the five qubit linear cluster state is identical to measuring $ZXZ$ for the three qubit cluster state while measuring  $ZZXZZ$ for the seventeen qubit cluster state is identical to measuring $ZZXZZ$ for the five qubit star cluster state. This shows that only the qubits that are actually measured influence the calculation of the visibility. As we can see in Fig.~(\ref{Fig:Vis}), as the initial cat state amplitude is increased, the visibility increases as well. It is also worth noting, that even if we had a perfect \csign gate for cat state logic, since the qubits are intrinsically non-orthogonal, the visibility is not automatically 1, as shown in Fig.~(\ref{Fig:Vis}).

\begin{figure}[ht]
\includegraphics[width=8cm]{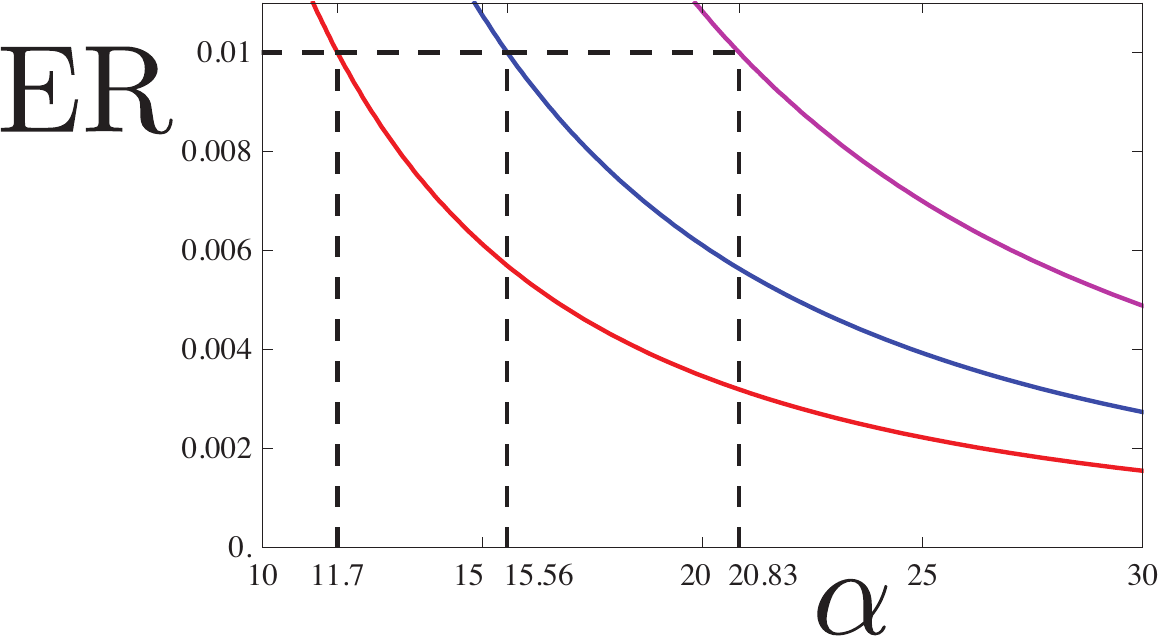}
\caption{\footnotesize The operator detection error rate per qubit (ER) for: detecting $XZ$ on the two qubit cluster state, shown in \textcolor{red}{red}; detecting $ZXZ$ on the five qubit linear cluster state, shown in \textcolor{blue}{blue}; detecting $ZZXZZ$ on the seventeen qubit cluster state, shown in \textcolor{magenta}{magenta}. The $\alpha$ values for each cluster state that gives an ER of $1\%$ are also shown. } 
\label{Fig:VisibilityERPlot}
\end{figure}

To judge how well our measurement model fits within the thresholds for the 3D topological cluster state code, we calculate the error rate per qubit for each of the visibilities in Fig.~(\ref{Fig:Vis}) 
by comparing with the visibilities resulting from measuring an ideally constructed qubit cluster state in which each qubit has undergone a depolarising error~\cite{NielsenBook}, as done in Section~\ref{sec:basicscheme}.
This is shown in Fig~(\ref{Fig:VisibilityERPlot}). The ER for the two qubit cluster is below $1\%$ when $\alpha>11.7$, while the ER for the five qubit linear cluster is below $1\%$ when $\alpha>15.56$ and the ER for the seventeen qubit cluster is below $1\%$ when $\alpha>20.83$. Since $XXZXX$ is a local stabiliser for the 3D topological code, to ballistically construct a 3D topological cluster state with coherent state logic, the amplitude for initial cat state would need to be greater than $\alpha=20.83$.

\section{\label{sec:teleportation}Telelportation}

\begin{figure}[ht]
\includegraphics[width=8cm]{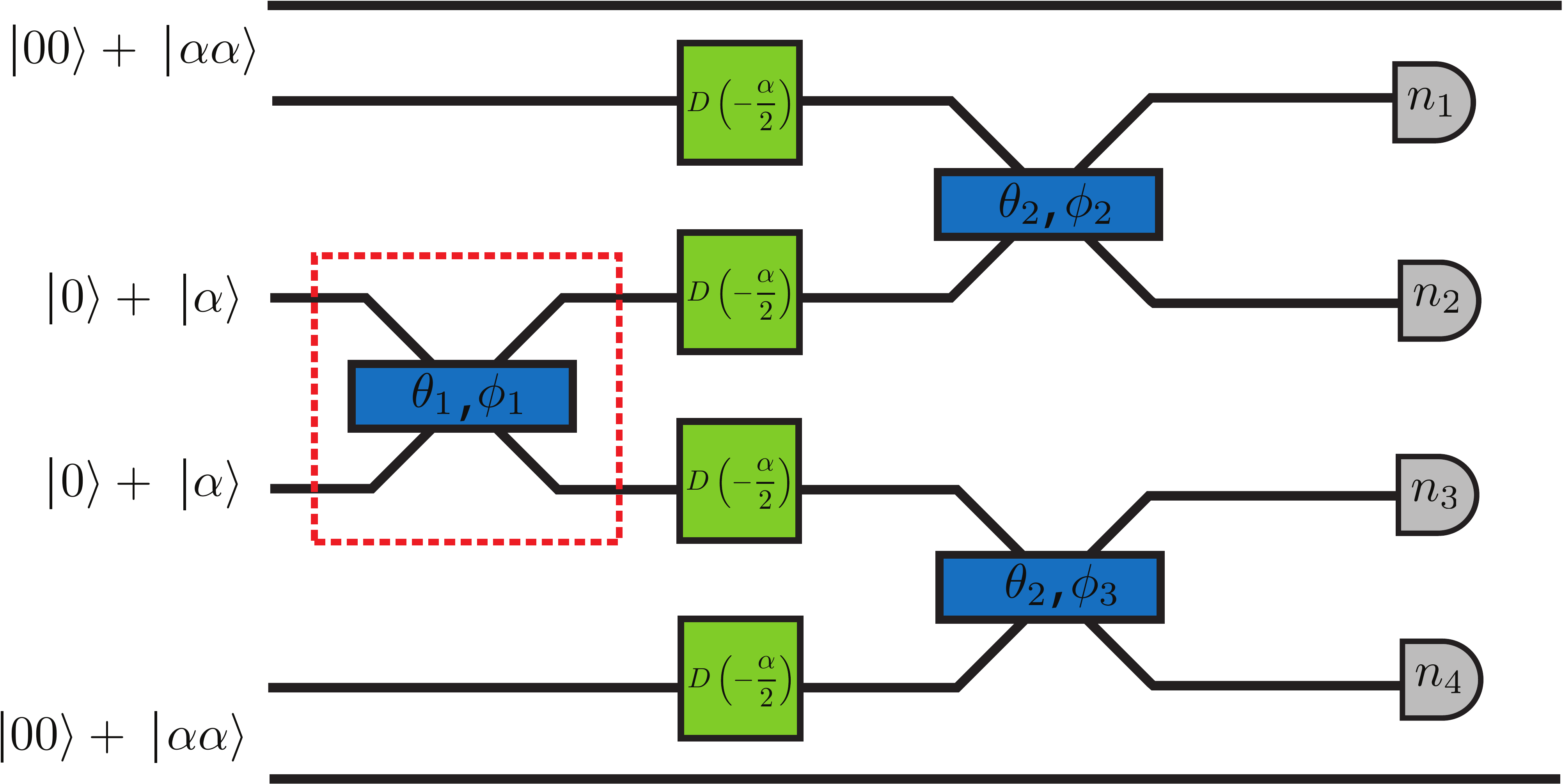}
\caption{\footnotesize Teleportation of the ballistic \csign gate (dashed box): $\theta_{1}=\frac{\pi}{2\alpha^{2}}$, $\phi_{1}=-\frac{\pi}{2}$, $\theta_{2}=\frac{\pi}{4}$, $\phi_{2}=0$, $\phi_{3}=\pi$.} 
\label{Fig:CleanTele}
\end{figure}

In this section we circumvent the low fidelities achieved from using a beam splitter as a \csign gate shown in Fig~(\ref{Fig:BasicFidelitiesPlot}) by attempting to {\em clean up} the cluster states by incorporating teleportation. We base our teleportation protocol, shown in Fig.~(\ref{Fig:CleanTele}), on a similar teleportation idea from Ralph~\etal~\cite{Ralph03a}. We teleport our basic two qubit cluster state using two copies of the Bell state $\ket{0,0}+\ket{\alpha,\alpha}$. The fact that we also need to generate this Bell state places a more stringent restriction on the size of initial cat states we need to implement \csign gates -- instead of just the state $\ket{0}+\ket{\alpha}$, we now also need a supply of the larger amplitude cat states $\ket{0} +\ket{\sqrt{2}\alpha}$, since the Bell state $\ket{00}+\ket{\alpha\alpha}$ is the output of $\ket{0} +\ket{\sqrt{2}\alpha}$ incident on one port of a symmetric beam splitter.

\begin{figure}[ht]
\includegraphics[width=6cm]{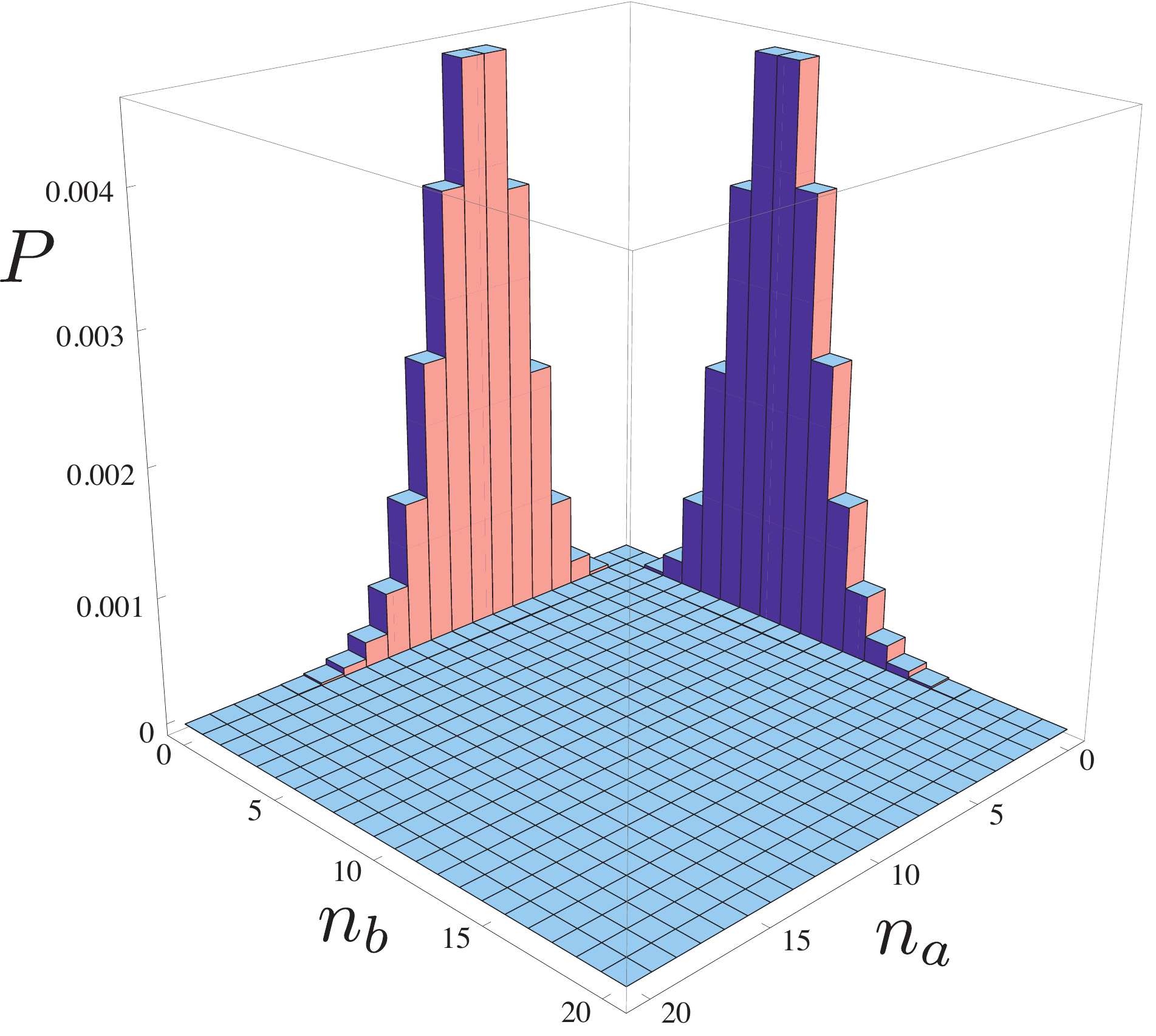}
\caption{\footnotesize The success probability as a function of the detection outcome for the teleporter in Fig.(\ref{Fig:CleanTele}) when $\alpha=\sqrt{2}\times 4\approx 5.66$. In this case we examine detector outcomes of the form $\{n_{1},n_{2},n_{3},n_{4}\}=\{n_{a},n_{b},n_{a},n_{b}\}$. } 
\label{Fig:TeleProbAlpha4p0NaNbNaNb}
\end{figure}

\begin{figure}[ht]
\includegraphics[width=6cm]{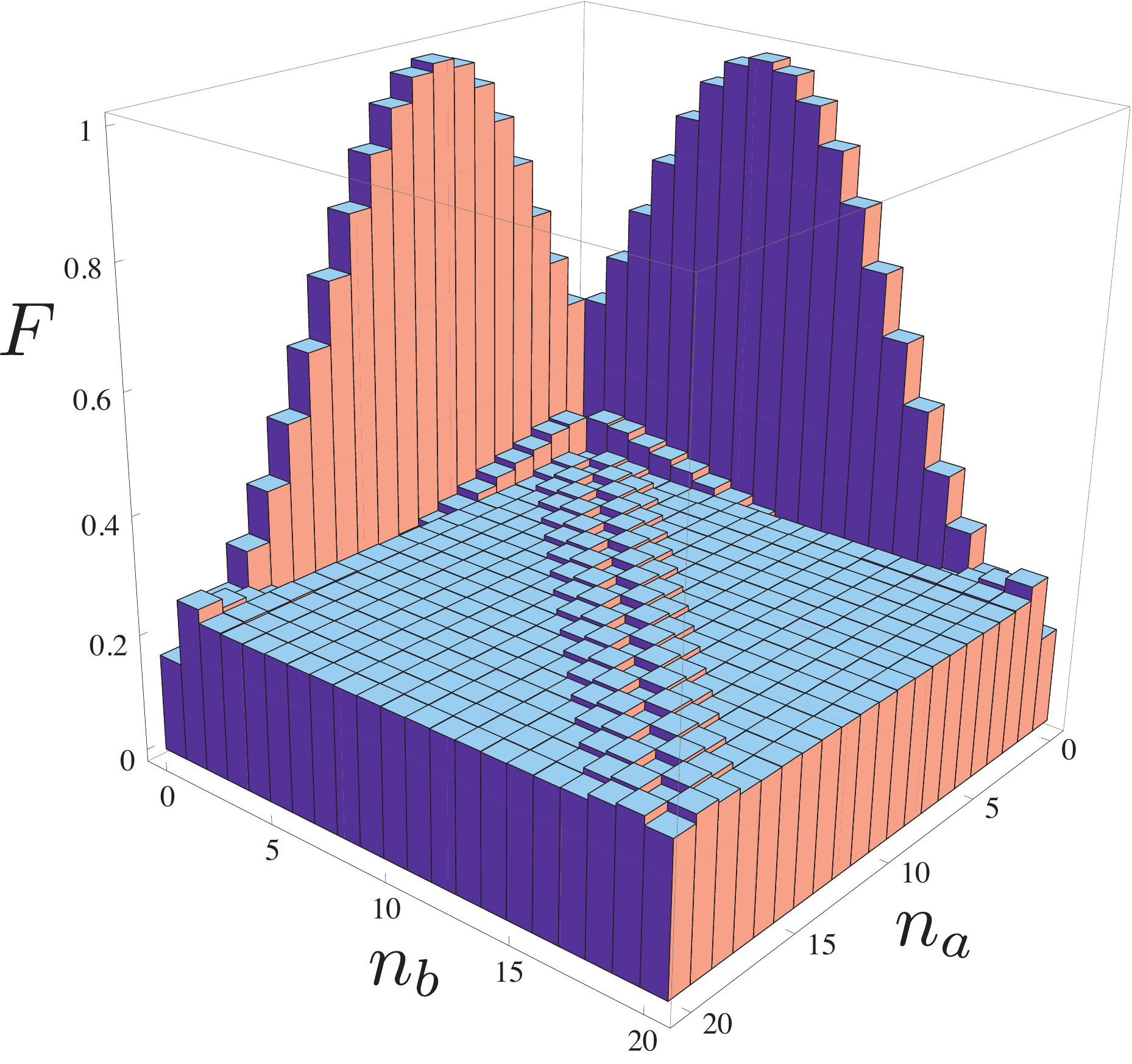}
\caption{\footnotesize The normalised fidelity as a function of the detection outcome for the teleporter in Fig.(\ref{Fig:CleanTele}) when $\alpha=\sqrt{2}\times 4\approx 5.66$. In this case we examine detector outcomes of the form $\{n_{1},n_{2},n_{3},n_{4}\}=\{n_{a},n_{b},n_{a},n_{b}\}$. } 
\label{Fig:TeleFidAlpha4p0NaNbNaNb}
\end{figure}

In Figs.~(\ref{Fig:TeleProbAlpha4p0NaNbNaNb}) and~(\ref{Fig:TeleFidAlpha4p0NaNbNaNb}) we examine the detection outcomes for this teleportation scheme. As can be seen from these figures and when one closely examines the output of the teleporter, there are four dominant detection sequences: 
\begin{align}
\{n_{1},n_{2},n_{3},n_{4}\} = \{n_{a},0,n_{b},0\}, &\{n_{a},0,0,n_{b}\}, \{0, n_{a},n_{b},0\},\nonumber\\
& \{0, n_{a},0,n_{b}\}\label{eqn:DetectorOutcomes}
\end{align}
We also notice in Figs.~(\ref{Fig:TeleProbAlpha4p0NaNbNaNb}) and~(\ref{Fig:TeleFidAlpha4p0NaNbNaNb})  that the maximum success probability and normalised fidelity is centred around $\alpha^{2}/2$, in this case around $\alpha=8$.

In contrast to~\cite{Ralph03a}, the fidelity in Fig.~(\ref{Fig:TeleFidAlpha4p0NaNbNaNb}) only reaches values close to one when the success probability in Fig.~(\ref{Fig:TeleProbAlpha4p0NaNbNaNb}) is a maximum. This is due to the delicacy of teleporting the phase information hidden in the approximate \csign state in Eq.~(\ref{eqn:actualCSstate})  -- any measurement outcome that is not one of the four sequences in Eq.~(\ref{eqn:DetectorOutcomes}) removes all phase information on the output, which effectively becomes the identity state. However, since the probability is close to 0 in the regions where the fidelity is less than ideal, this does not have a dominant affect on the success of the teleporter.  

\begin{figure}[ht]
\includegraphics[width=6cm]{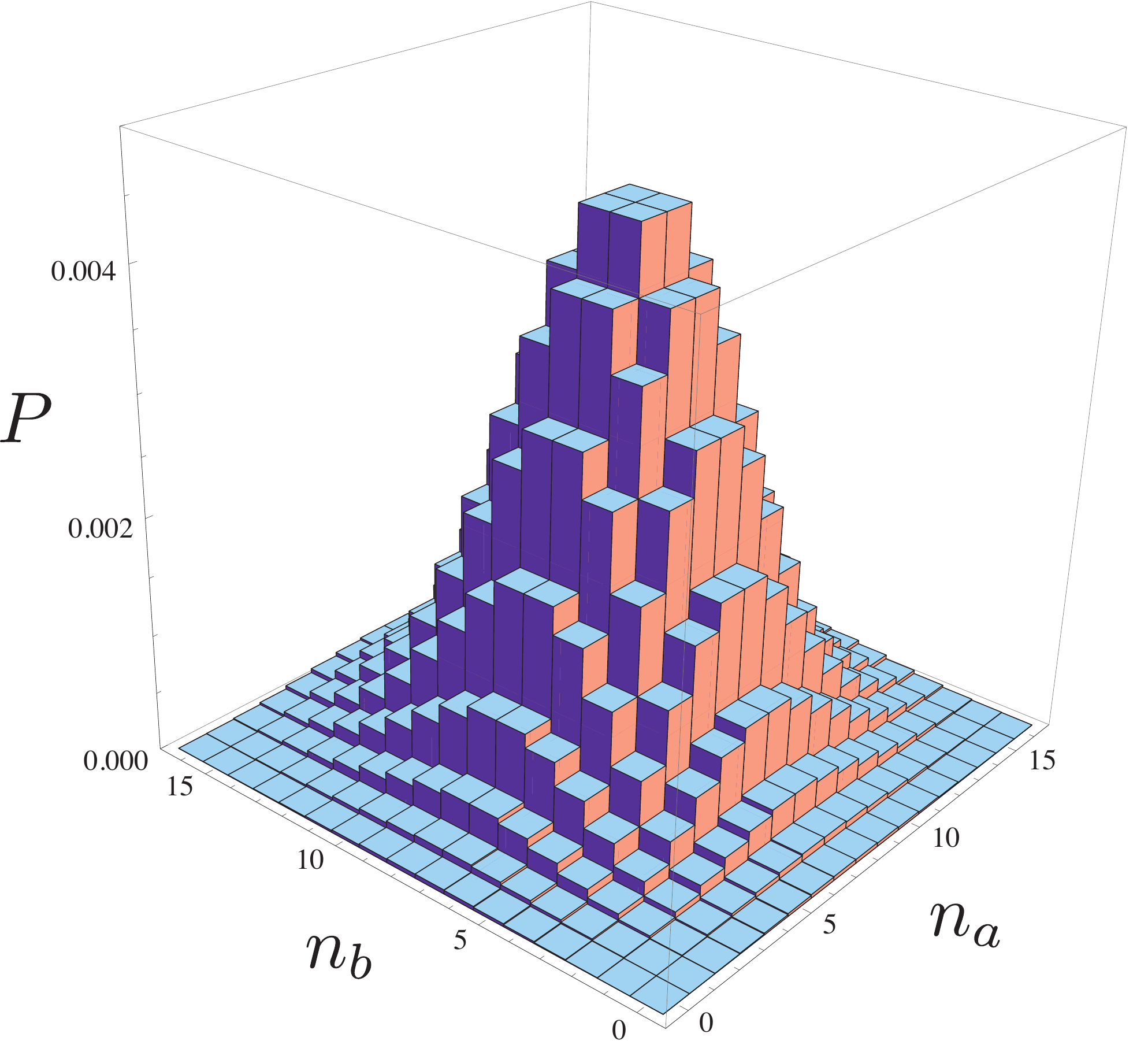}
\caption{\footnotesize The success probability as a function of the detection outcome for the teleporter in Fig.(\ref{Fig:CleanTele}) when $\alpha=\sqrt{2}\times 4\approx 5.66$. In this case we examine detector outcomes of the form $\{n_{1},n_{2},n_{3},n_{4}\}=\{n_{a},0 ,0 ,n_{b}\}$.} 
\label{Fig:TeleProbAlpha4p0Na00Nb}
\end{figure}

\begin{figure}[ht]
\includegraphics[width=6cm]{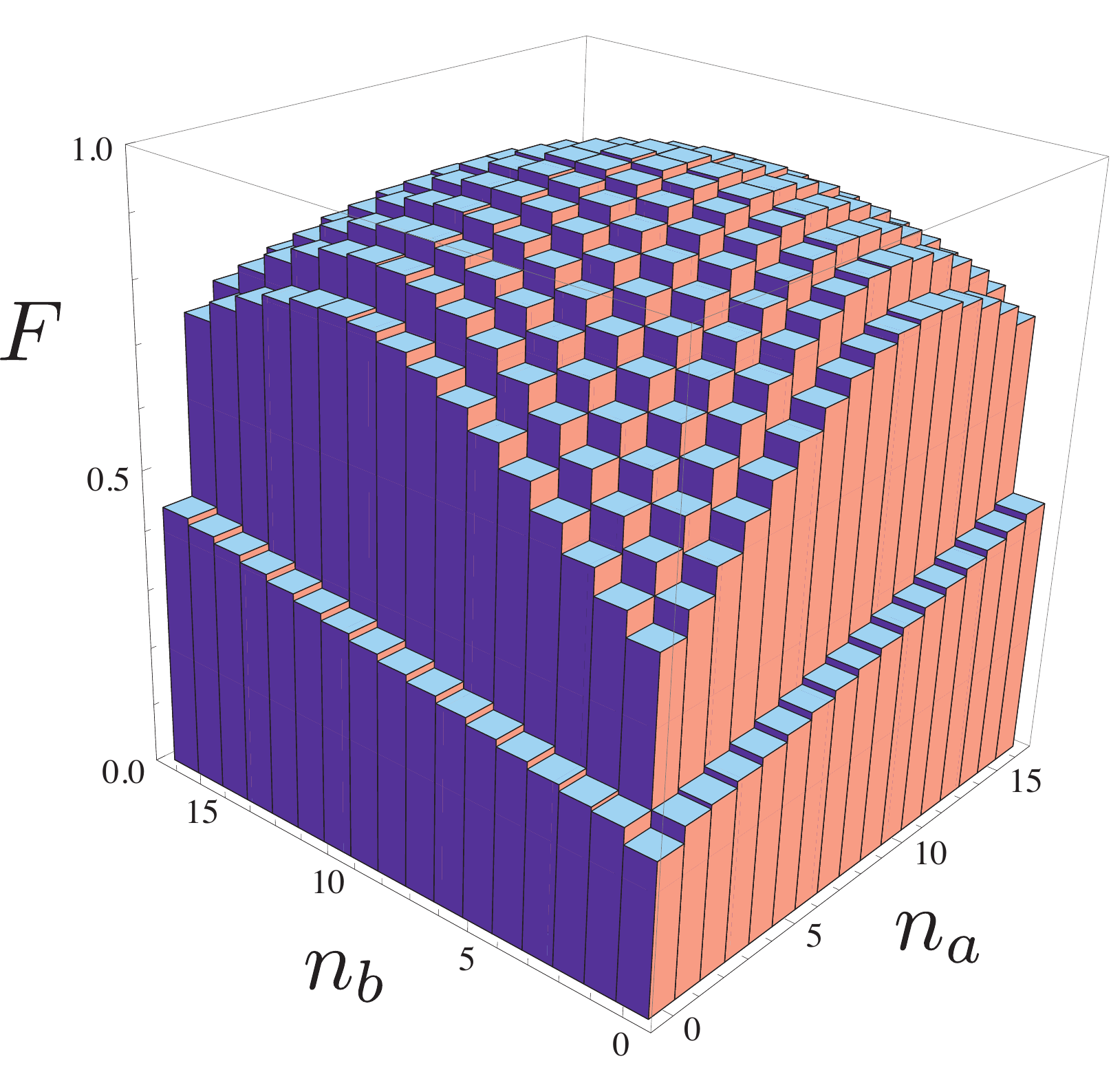}
\caption{\footnotesize The normalised fidelity as a function of the detection outcome for the teleporter in Fig.(\ref{Fig:CleanTele}) when $\alpha=\sqrt{2}\times 4\approx 5.66$. In this case we examine detector outcomes of the form $\{n_{1},n_{2},n_{3},n_{4}\}=\{n_{a},0 ,0 ,n_{b}\}$.} 
\label{Fig:TeleFidAlpha4p0Na00Nb}
\end{figure}

Given that we accept the teleporter detections are of the form given in Eq.~(\ref{eqn:DetectorOutcomes}), we can examine the success probability and normalised fidelity as a function of $n_{a}$ and $n_{b}$, as shown in Figs.~(\ref{Fig:TeleProbAlpha4p0Na00Nb}) and~(\ref{Fig:TeleFidAlpha4p0Na00Nb}). We again notice that the normalised fidilty in Fig.~(\ref{Fig:TeleFidAlpha4p0Na00Nb}) is only close to 1 when the success probability in Fig.~(\ref{Fig:TeleProbAlpha4p0Na00Nb}) is a maximum.

As with standard qubit teleportation~\cite{Bennett93}, there are single qubit corrections necessary on the output of Fig.~(\ref{Fig:CleanTele}). We make the assumption that these corrections can be performed at a later time, or we could explicitly avoid applying these corrections altogether by keeping track of the necessary corrections, staying in the so called Pauli frame~\cite{Knill05a}, compensating for these corrections in subsequent measurements. 

\begin{figure}[ht]
\includegraphics[width=8cm]{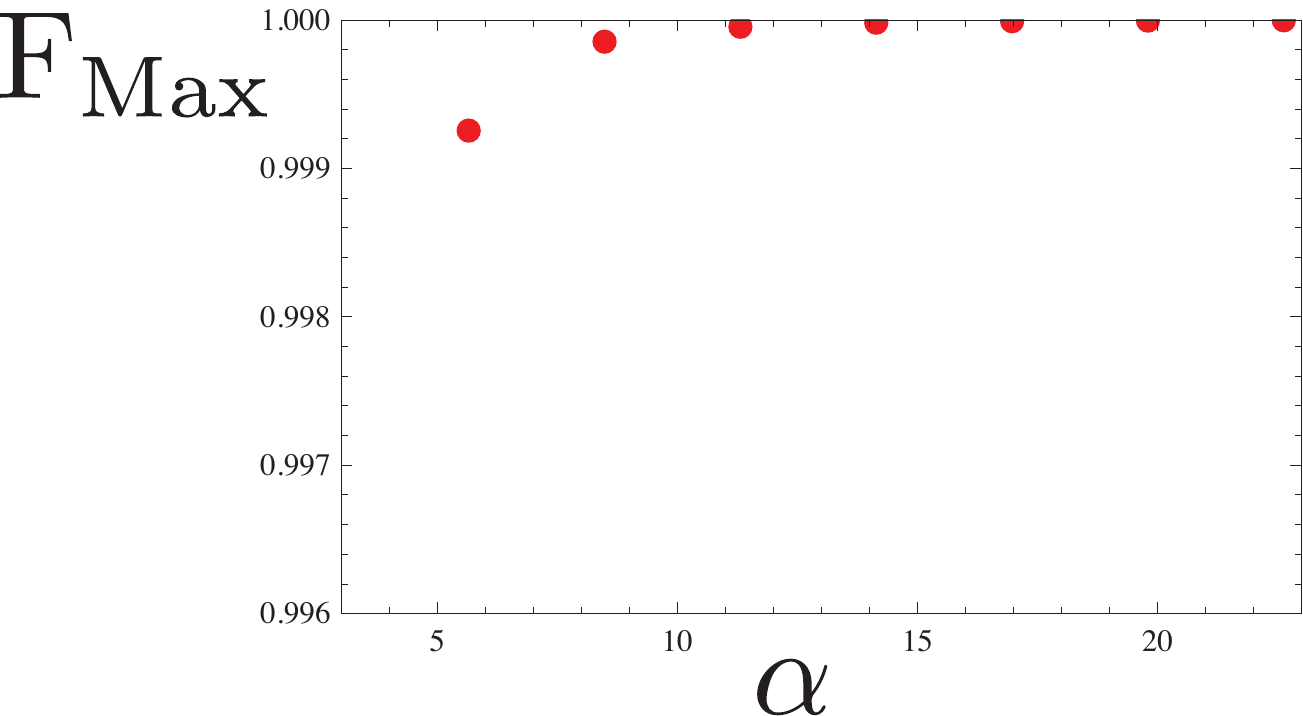}
\caption{\footnotesize The maximum fidelity the \csign teleporter in Fig~(\ref{Fig:CleanTele}) can produce as a function of $\alpha$. Note that $\alpha$ has been scaled to take the preparation of the Bell state cat states into account, that is, $\alpha\to \sqrt{2}\alpha$.} 
\label{Fig:MaxFid2Qtele}
\end{figure}

\begin{figure}[ht]
\includegraphics[width=8cm]{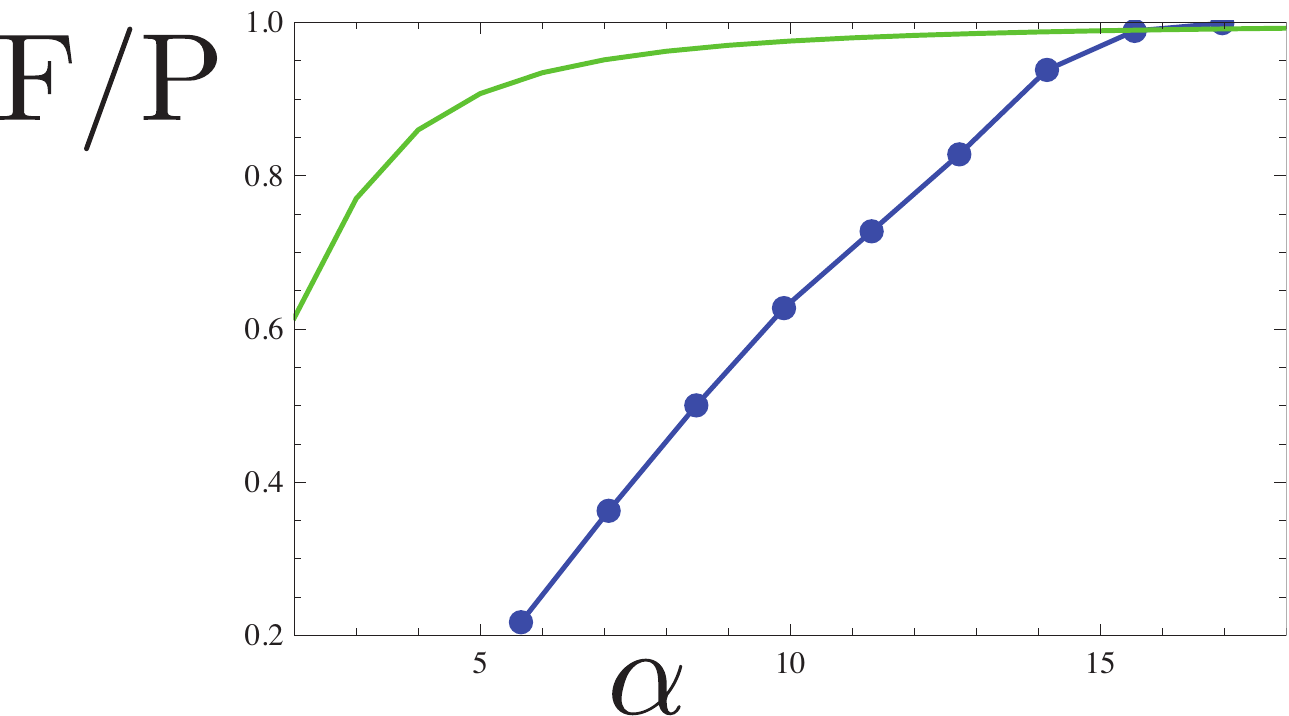}
\caption{\footnotesize The success probability of the teleporter in Fig.~(\ref{Fig:CleanTele}) given that the output must have an average fidelity $>99\%$ is shown in \textcolor{blue}{blue} and the fidelity for the beam splitter \csign gate from Fig.~(\ref{Fig:CSignBS}) is shown in \textcolor{green}{green}.  Note that $\alpha$ has been scaled for the teleporter probability curve to take the preparation of the Bell state cat states into account, that is, $\alpha\to \sqrt{2}\alpha$.} 
\label{Fig:TeleFidAndProgFavGr99Ptot}
\end{figure}

The normalised fidelity of the of the teleporter in Fig.~(\ref{Fig:CleanTele}) can be made arbitrarily close to 1, given the correct detector firing. For an initial cat state amplitude as low as $\sqrt{2}\times 2\approx 2.83$, the maximum normalised fidelity can be made as high as $0.999$. This is shown in Fig.~(\ref{Fig:MaxFid2Qtele}). The trade-off for having a high fidelity is the introduction of a success probability. We can see how the success probability of the teleporter scales given that we demand a certain average fidelity for the output. The average fidelity is given by~\cite{Ralph03a}
\begin{align}
\text{F}_{\text{av}}=\frac{1}{P_{\text{det}}}\sum_{(n_{a},n_{b})\in S}P(n_{a},n_{b})F(n_{a},n_{b}),
\end{align}
where $P_{det}=\sum_{(n_{a},n_{b})\in S}P(n_{a},n_{b})$ is the probability that the detection outcomes are in the set $S$ and $F(n_{a},n_{b})$ is the fidelity between the normalised output of Fig.~(\ref{Fig:CleanTele}) and Eq.~(\ref{eqn:idealCSstate}). In Fig.~(\ref{Fig:TeleFidAndProgFavGr99Ptot}), we calculate the success probability of the teleporter given that we demand the output must have an average fidelity $>99\%$. As can be seen, for $\alpha$ large enough, the success probability  approaches 1. 

In the next section we will examine whether teleporting the the basic \csign state is advantageous, that is, whether using a teleporter can reduce the amplitude of the initial cat states.

\section{\label{sec:comparison} Discussion}

In this section we discuss the performance of incorporating teleportation into our 3D cluster state production. To determine this we examine whether introducing located loss errors through teleportation can reduce the computational basis errors to a level that is correctable by the 3D topological cluster state code~\cite{Raussendorf07a, Raussendorf07b}. We consider two possible simultaneous computational basis error and detectable loss error thresholds for the  3D topological cluster state code. We present preliminary results in which we consider the teleportation of a two qubit cluster. The teleportation of larger cluster states will be the subject of a future paper. 

We first consider the most stringent and thorough threshold, predicted by Barrett and Stace~\cite{Barrett10a}, in which it is estimated 3D topological cluster states could correct for sole computational basis errors as high as $0.63\%$ and sole located loss errors as high as $24.9\%$, with a close to linear relationship for simultaneous loss and computation errors between these two bounds. We next consider a more optimistic outlook for the capabilities of these topological codes, by relaxing the computation basis error bound to $1\%$, a value that is believed could be reached~\cite{Fowler09a}, and keeping the $24.9\%$ located loss error bound.

\begin{figure}[ht]
\includegraphics[width=8.6cm]{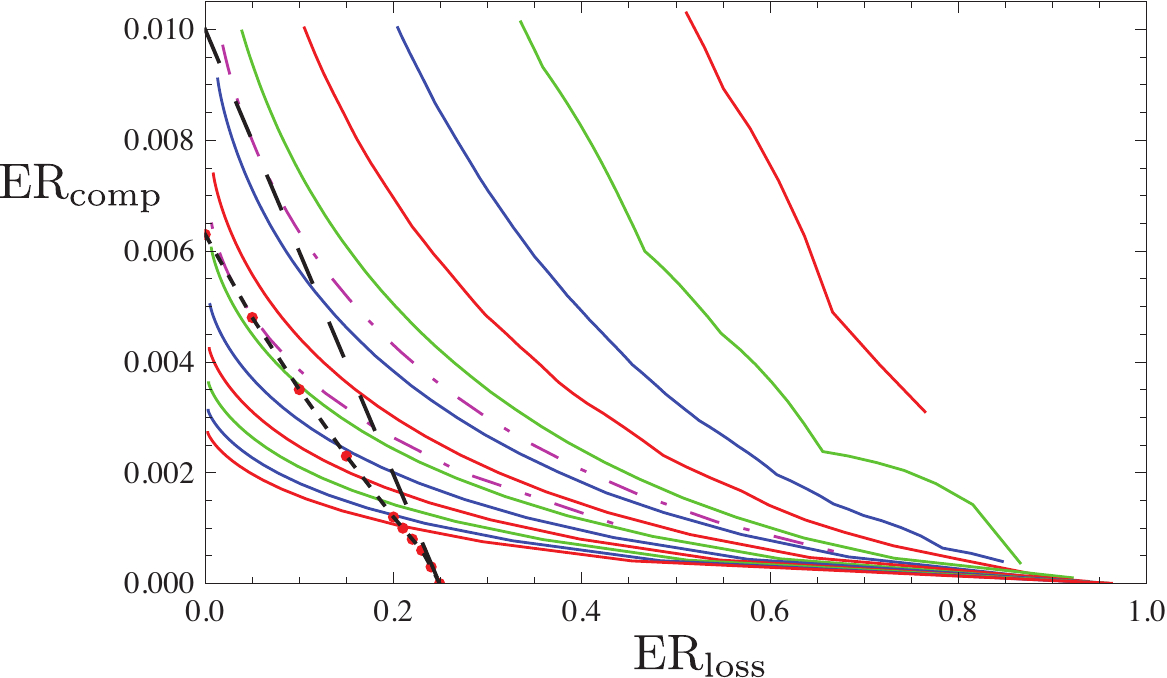}
\caption{\footnotesize Graph showing the relationship between computational basis and located loss error rates for various initial cat state amplitudes. The dotted black line on the left shows the predicted threshold by Barrett and Stace~\cite{Barrett10a}. The dashed black line on the right shows the more optimistic threshold that encompasses the Barrett and Stace located loss error rate and the believed best case computation loss error rate~\cite{Fowler09a}. The initial cat state amplitudes increase from right to left. The solid lines correspond to $\alpha=\sqrt{2}\times(3,4,5,6,7,8,9,10,11,12,13,14,15)$ $=(4.24, 5.66, 7.07, 8.49, 9.90,$  $11.31, 12.73, 14.14, 15.56,$ $16.97, 18.38, 19.80, 21.21)$ (from right to left). The dot-dashed line on the left corresponds to $\alpha=\sqrt{2}\times 9.65=13.65$ and the dot-dashed line on the right corresponds to $\alpha=\sqrt{2}\times 7.56=10.69$.} 
\label{Fig:ERLossCompAll}
\end{figure}

In Fig~(\ref{Fig:ERLossCompAll}), we examine the relationship between computational basis and located loss error rates for teleportation. Before teleportation, each $\alpha$ has an intrinsic computation error rate associated with it, as shown in Fig.~(\ref{Fig:BasicFidelitiesPlotER}). The aim of this exercise to determine whether it is worthwhile to teleport the state in Eq.~(\ref{eqn:actualCSstate}). This is done by keeping track of the detections outcomes from the teleporter that give the highest output fidelity. By counting the detection outcomes, we generate the curves shown in Fig.~(\ref{Fig:ERLossCompAll}). 

For example, consider the rightmost curve, corresponding to $\alpha=\sqrt{2}\times 3.0=4.24$. Initially we only consider a few detection outcomes centred around $(\alpha/2)^{2}$, leading to a small teleportation success probability, which means a large located loss error. However, since these detections correspond to high fidelity outputs, the computational error rate is low, which is why the curve begins at the bottom right of the graph. As we consider more detection outcomes, the teleportation success probability increases, reducing the located loss error rate, at the expense of allowing more low fidelity output states, therefore reducing the computational basis error rate. Consequently, the $\alpha=4.24$ curve moves from the bottom right to the top left. 

The number of detection outcomes considered for each curve in Fig.~(\ref{Fig:ERLossCompAll}) increased as function of $\alpha$. For low values of $\alpha$, there are not many teleportation detection outcomes that result in a computational basis error rate less that $1\%$, which is why these initial curves are not smooth.

The leftmost dot-dashed curve in Fig.~(\ref{Fig:ERLossCompAll}) shows that teleporting coherent cat states with an amplitude of $\alpha=\sqrt{2}\times 9.65=13.65$ intersects the Barrett-Stace~\cite{Barrett10a} threshold. The no located loss threshold for this case is equivalent to a computation basis error rate of $0.63\%$, which corresponds to the output from the basic \csign gate in Fig.~(\ref{Fig:CSignBS}) with $\alpha=13.96$. This means teleportation would allow initial cat states with $(13.96/2)^{2}-(13.65/2)^{2}=2.16$ less photons. These result are summarised in the second column of Tab.~(\ref{table:TeleVsBallisticSumm}).

The rightmost dot-dashed curve in Fig.~(\ref{Fig:ERLossCompAll}) shows that teleporting coherent cat states with an amplitude of $\alpha=\sqrt{2}\times 7.56=10.69$ intersects the more optimistic threshold. The no located loss threshold for this case is equivalent to a computation basis error rate of $1.0\%$, which corresponds to the output from the basic \csign gate in Fig.~(\ref{Fig:CSignBS}) with $\alpha=11.07$. This means teleportation would allow initial cat states with $(11.07/2)^{2}-(10.69/2)^{2}=2.06$ less photons. These result are summarised in the third column of Tab.~(\ref{table:TeleVsBallisticSumm}).

\begin{table}[h]
\begin{center}
\begin{tabular}{|c|c|c|}
\hline
 & Barrett & Opt.  \\
\hline
\hline
$\text{ER}_{\text{loss}}$ &$24.9\%$&$24.9\%$\\
\hline
$\text{ER}_{\text{comp}}$ &$0.63\%$&$1\%$\\
\hline
Basic \csign $\alpha$ &13.96&11.07\\
that gives $\text{ER}_{\text{comp}}$ &&\\
\hline
Teleportation $\alpha$ &$\sqrt{2}\times 9.65$&$\sqrt{2}\times 7.56$\\
&$=13.65$&$=10.69$\\
\hline
$\text{ER}_{\text{comp}}$ for this $\alpha$  &$0.66\%$&$1.07\%$\\
\hline
Number of photons &46.56&28.58\\
\hline
Photon $\#$ reduction & 2.16& 2.06\\
achieved by teleporting &&\\
\hline
\end{tabular}
\caption{\footnotesize A summary of teleportation vs ballistic \csign gates for each of the three threshold cases considered. }
\label{table:TeleVsBallisticSumm}
\end{center}
\end{table} 

The results from Fig.~(\ref{Fig:ERLossCompAll}) that are summarised in Tab.~(\ref{table:TeleVsBallisticSumm}) suggest that depending on which threshold is ultimately correct for 3D topological cluster state codes, teleportation will always result in less stringent restriction being put on the amplitude size of the initial cat state, albeit in some cases only a reduction of $2.06$ photons on average. However, it should be noted that a full analysis for the teleportation of multi-qubit cluster state is necessary before we can claim that teleportation allows for a reduction in coherent amplitude when compared to the completely ballistic scheme. 

It is worth exploring the possibility of not being penalised for using Bell states of the form $\ket{00}+\ket{\alpha\alpha}$ for teleportation. If this was conceivable, then teleportation would offer much greater benefits, allowing initial cat states with up to $25$ less photons on average. The results from Tab.~(\ref{table:TeleVsBallisticSumm}) are re-analysed for such a case in Tab.~(\ref{table:TeleVsBallisticSummNoPenalty}). Possible ways of avoiding these penalties would be to have a reliable source of $\ket{00}+\ket{\alpha\alpha}$ states, such as the entangled cat state produced by Ourjoumtsev~\etal~\cite{Ourjoumtsev09a}. However, we could then consider the more complicated entangled cat states required for quantum gate teleportation in Lund~\etal~\cite{Lund08a} and proceed with cluster state production purely via teleportation.

\begin{table}[h]
\begin{center}
\begin{tabular}{|c|c|c|}
\hline
 & Barrett & Opt.  \\
\hline
\hline
$\text{ER}_{\text{loss}}$ &$24.9\%$&$24.9\%$\\
\hline
$\text{ER}_{\text{comp}}$ &$0.63\%$&$1\%$\\
\hline
Basic \csign $\alpha$ &13.96&11.07\\
that gives $\text{ER}_{\text{comp}}$ &&\\
\hline
Teleportation $\alpha$ &$9.65$&$ 7.56$\\
\hline
$\text{ER}_{\text{comp}}$ for this $\alpha$  &$1.31\%$&$2.13\%$\\
\hline
Number of photons &23.28&14.29\\
\hline
Photon $\#$ reduction &25.44&16.35\\
achieved by teleporting &&\\
\hline
\end{tabular}
\caption{\footnotesize A summary of teleportation vs ballistic \csign gates for each of the three threshold cases considered, assuming we did not have to take the construction of the Bell state $\ket{00}+\ket{\alpha \alpha}$ into account for teleportation.}
\label{table:TeleVsBallisticSummNoPenalty}
\end{center}
\end{table}

\section{\label{sec:conclusions}Conclusion}

We have shown that the ballistic construction of 3D topological cluster states with coherent logic qubits is possible provided the initial cat states are large enough. Since the entangling gates for this scheme only involve passive linear optical elements, the use of integrated quantum optical circuits would be ideally suited to implement this scheme. We found that cat states with an average number of photons $>108$ ($\ket{\alpha>9.25}+\ket{\alpha<-9.25}$) gave an error rate per qubit for the five qubit star cluster state less than $1\%$, a computational error fault tolerant threshold that is believed the 3D topological cluster states will satisfy. 

We have also shown that teleportation could be used to {\em clean up} the cluster states produced from ballistic gates, and that teleportation might provide a method to trade-off computational basis errors for located loss errors. Preliminary results for the teleportation of two qubit cluster states suggests that initial cat states with an amplitude of $\alpha=\sqrt{2}\times7.56=10.69$ (an average photon number of $28.58$) would result in a combined located loss error rate and computational basis error rate that is below an optimistic threshold for 3D topological cluster state codes.  

In this paper we have assumed that all errors arise from the  3D cluster state construction in the form of computational basis errors via the ballistic \csign gate and located loss errors via teleportations to clean up the ballistic \csign gate. In reality there will be other sources of error, such as measurement errors, errors resulting from actual computation and storage errors.
In addition to this, the first physical requirement for topological cluster state computation, the need for state preparations of $\ket{0}_{L}$ and $\ket{0}_{L}+e^{i\theta}\ket{1}_{L}$, still needs to be addressed for coherent state logic. 

\begin{acknowledgments}

We would like to thank Austin Fowler, Simon Devitt, Kae Nemoto and Bill Munro for valuable discussions. This work was supported by the Australian Research Council.

\end{acknowledgments}



\begin{thebibliography}{99}

\bibitem{Raussendorf07a} R.~ Raussendorf and J.~Harrington, Phys. Rev. Lett. {\bf 98}, 190504 (2007).

\bibitem{Raussendorf07b} R.~ Raussendorf, J.~Harrington and K.~Goyal, New J. Phys. {\bf 9}, 199 (2007).

\bibitem{Fowler09a} A.G.~Fowler and K.~Goyal, Quant. Info. Comput. {\bf 9}, 721 (2009).

\bibitem{Devitt09a} S.J.~Devitt~\etal, New J. Phys. {\bf 11}, 083032 (2009).

\bibitem{Barrett10a} S.D.~Barrett and T.M.~Stace, Phys. Rev. Lett. {\bf 105}, 200502 (2010).

\bibitem{KLM} E. Knill, R. Laflamme and G.J. Milburn, Nature  {\bf 409}, 46 (2001).

\bibitem{Ralph02a} T.~C.~Ralph, W.~J.~Munro and G.~J.~Milburn, In Proc. SPIE {\bf 4917}, 1 (2002).

\bibitem{Ralph03a} T.~C.~Ralph~\etal, Phys. Rev. A {\bf 68}, 042319 (2003).

\bibitem{Politi08a} A.~Politi~\etal, Science {\bf 320}, 646 (2008).

\bibitem{Matthews09a} J.C.F.~Matthews~\etal, Nature Phot. {\bf 3}, 346 (2009).

\bibitem{Marshall09a} G.D.~Marshall~\etal, Opt. Exp. {\bf 17}, 12546 (2009).

\bibitem{Smith09a} B.J.~Smith~\etal, Opt. Exp. {\bf 17}, 13516 (2009).

\bibitem{Takahashi08a} H.~Takahashi~\etal, Phys. Rev. Lett. {\bf 101}, 233605 (2008).

\bibitem{Gerrits10a} T.~ Gerrits~\etal, Phys. Rev. A {\bf 82}, 031802(R) (2010).

\bibitem{Lund08a} A.P.~Lund, T.C.~Ralph and H.L.~Haselgrove, Phys. Rev. Lett. {\bf 100}, 030503 (2008).

\bibitem{WallsBook} D.F.~Walls and G.J.~Milburn, \textit{Quantum Optics}, Springer-Verlag, Berlin, (1994).

\bibitem{NielsenBook} M.A.~Nielsen and I.L.~Chuang, \textit{Quantum Computing and Quantum Information}, (Cambridge University Press) 2000.

\bibitem{GardinerBook} C.W.~Gardiner and P.~Zoller., \textit{Quantum Noise}, Springer-Verlag, Berlin, (2000).

\bibitem{AbramowitzBook} M.~Abramowitz and I.A.~ Stegun., \textit{Handbook of Mathematical Functions}, Dover Publications, New York, (1972).

\bibitem{Bennett93} C.H.~Bennett, {\em et~al.},  Phys. Rev. Lett. {\bf 70}, 1895  (1993).

\bibitem{Knill05a} E.~Knill, Phys. Rev. A {\bf 71}, 042322 (2005).

\bibitem{Ourjoumtsev09a} A.~Ourjoumtsev~\etal, Nat. Phys. {\bf 5}, 189 (2009).

\end{thebibliography}
\end{document}